\documentclass{aa}
\usepackage{natbib} 
\usepackage{graphicx}
\usepackage{color}
\usepackage{times}
\usepackage{booktabs}
\usepackage{txfonts}

\begin{document}

\title{The challenge of large and empty voids in the SDSS DR7 redshift survey}

\author{Saeed Tavasoli\inst{1,2}
 ,  Kaveh Vasei\inst{2,3} and Roya Mohayaee\inst{1}}
\institute{
$^1$Institut d$^{'}$Astrophysique de Paris (IAP), CNRS, UPMC, 98 bis boulevard Arago, France\\
$^2$Institute for Research in Fundamental Sciences (IPM), P.O. Box 19395-5531, Tehran, Iran \\
$^3$Department of Physics and Astronomy, University of California, Riverside, CA 92521, USA 
}

\titlerunning{Voids in SDSS DR7 survey}
\authorrunning{Tavasoli, Vasei, Mohayaee}
\date{Received 21 Novembre 2012}

% \abstract{}{}{}{}{} 
% 5 {} token are mandatory
 
  \abstract
   {
We present catalogues of voids for the SDSS DR7 redshift survey and
for Millennium I simulation mock data.
}
   {
We aim to compare the observations with simulations based on a $\Lambda$CDM model 
and a semi-analytic galaxy formation model.
We use the void statistics as a test for these models.
}
   {
We assembled a mock catalogue that closely resembles the 
SDSS DR7 catalogue and carried out a parallel statistical analysis of the observed and simulated catalogue.
}
   {
  We find that  
in the observation and the simulation, voids tend to be equally
spherical.  The total volume occupied by the voids and 
their total number are slightly larger in the simulation than in the observation.
We find that large voids are less abundant in the simulation
and the total luminosity of the galaxies contained in a void with a
given radius is higher on average than
observed by SDSS DR7 survey. 
We expect these discrepancies to be even more significant in reality than found here since
the present value of $\sigma_8$ given by
WMAP7 is lower than the value of $0.9$ used in the Millennium I simulation.
}
{The reason why the simulation fails to produce enough large and
dark voids might be the failure of certain semi-analytic galaxy
formation models to reduce the small-scale  power of $\Lambda$CDM
and to produce sufficient power on large scales.}


   \maketitle

\def\be{\begin{equation}}
\def\ee{\end{equation}}
\def\apj{{ApJ}}
\def\apjs{{ApJS}}
\def\apjl{{ApJL}}
\def\aap{{A\&A}}
\def\pasa{{PASA}}
\def\mnras{{MNRAS}}
\def\araa{{ARA\&A}}
\def\aj{{AJ}}
\def\prd{{PRD}}
\def\nat{{Nature}}

\section{Introduction}
\label{sec:introduction}

Redshift surveys have been demonstrating for several decades that galaxies are
distributed on a cosmic web of filaments, walls, and clumps. These
structures, which form on a hierarchy of scales and span a wide
redshift range, border low-luminosity regions that are mostly
devoid of observable galaxies. These ``void'' regions occupy more
than $80 \%$ of the volume of the observable Universe. Since the
discovery of voids using Zwicky clusters \citep{1980MNRAS.193.353E} and
the discovery of the first giant or supervoid in the Bootes
constellation \citep{1981ApJ...248L..57K} numerous works have
followed \citep{1982Natur.300..407Z,1982ApJ...253..423D,
1986ApJ...302L...1D,1988ApJ...327..544D,1989Sci...246..897G,1994ApJ...424L...1D} and diverse
algorithms for void identification have been developed and applied
to larger and more complete surveys 
 (see \citet{2008MNRAS.387..933C} for a summary and comparison of different methods).

\begin{figure*}
\center{
\includegraphics[scale=0.3,bb=0 30 700 600]{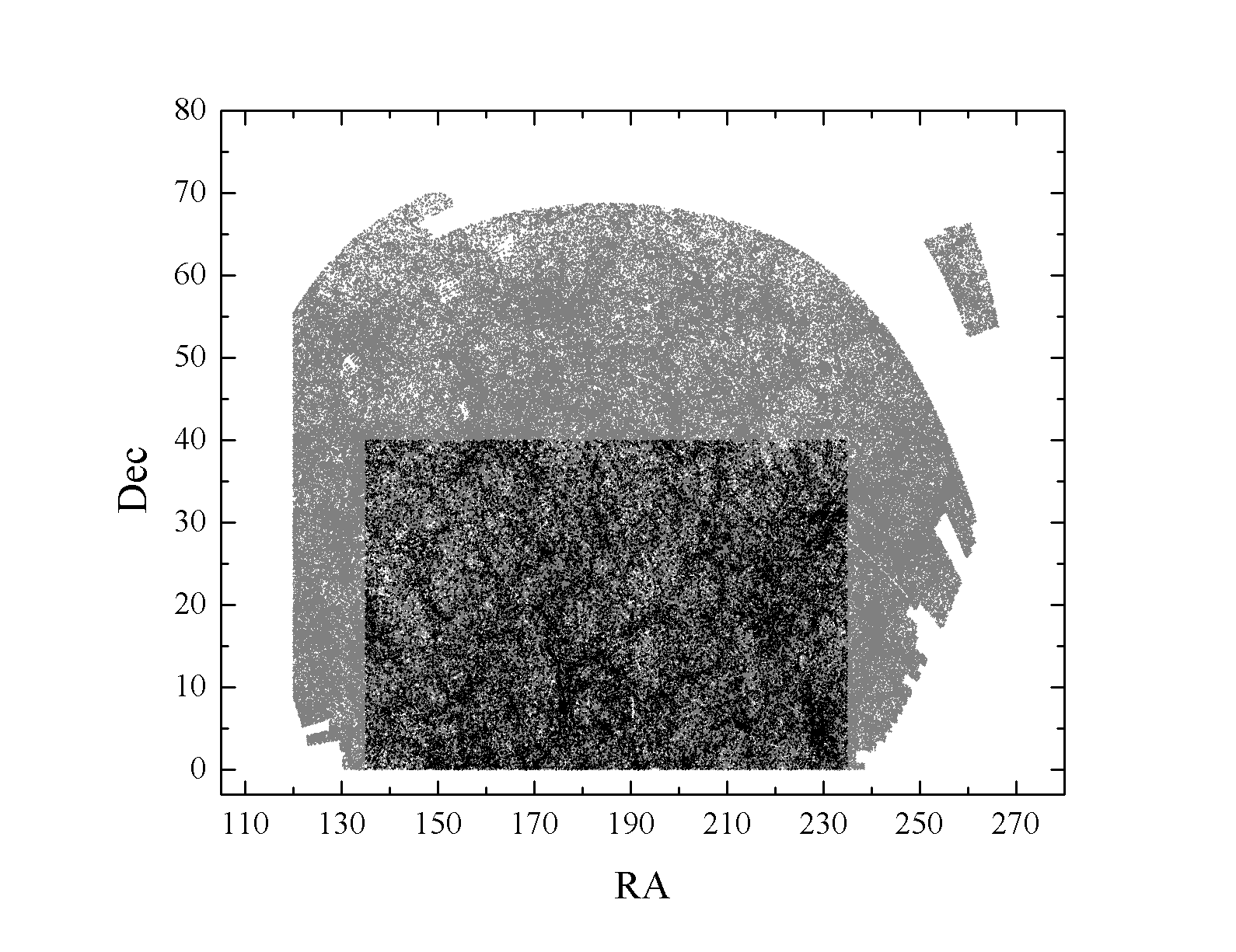}\hspace*{1cm}
\includegraphics[scale=0.3,bb=0 30 700 600]{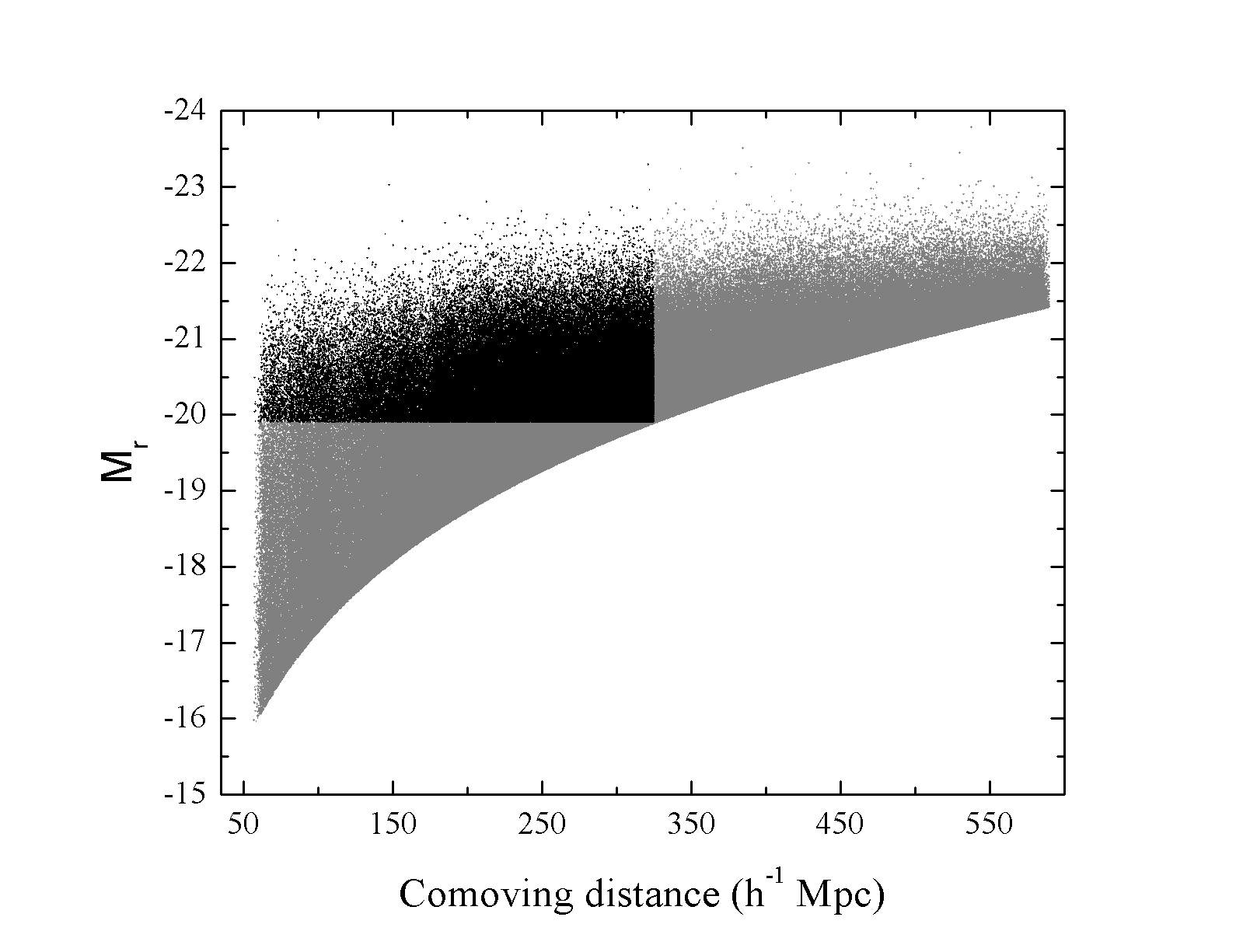}
\caption {Darker shaded areas in the two panels show the SDSS DR7 region (left) 
and the volume-limited sample (right) that we selected for this work.}}
\label{radec}
\end{figure*}

The formation and evolution of voids is well-understood in the
framework of gravitational instability \citep{1982Natur.300..407Z, 1989RvMP...61..185S}. 
However,  when one compares void properties of observations and simulations based on 
$\Lambda$CDM, certain problems still remain to be better understood. 
By definition, voids are devoid of galaxies
or contain only a negligible number of faint galaxies. The
perplexing issue is that we do not see a large population of
low-mass galaxies populating voids (\citet{1999mnras...522..82K}; \citet{1999ApJ.524.19M}),
 and furthermore, the void galaxies that we do see are basically
representative of the general population \citep{2001ApJ...557..495P}. 

Observed voids seem to contain fewer galaxies and in particular
dwarf galaxies, contrary to what is expected from $\Lambda$CDM
\citep{2001ApJ...557..495P, 2008ApJ...676..184T, 2009MNRAS.395.1915T}. 
Some studies have also shown that voids in observations 
are significantly larger than those in simulations  \citep{1984ApJ...287L..59R}.
Although modifying models of galaxy formation might solve these problems and various
remedies such as proper biasing and halo occupation distribution
have been proposed \citep{2005ApJ...620..618H,2008ApJ...686...53T}, different studies suggest that the
problem would still persist 
\citep{1986ApJ...308..510B,1994MNRAS.267..605L,2002MNRAS.330..399P,
2003MNRAS.344..715G,2004ApJ...607..751H,2005ApJ...621..643G,2006MNRAS.371..401H}. 

The problem of empty and large voids could arise because the
$\Lambda$CDM has too much power on small scales which would in turn lead to the
problem of over-abundance of substructures \citep{2009MNRAS.395.1915T}. 
Substructures would occupy the voids, making them less
empty, and statistically, they could break larger voids into smaller
ones. 
On the other hand, one could equally infer that  $\Lambda$CDM
lacks power on large scales, perhaps because the value of $\sigma_8$ is too low. 

In this work, we study this problem by analysing voids in the SDSS DR7 data and 
by carrying out a parallel and comparative analysis on a mock-SDSS DR7 catalogue based on the Millennium I simulation.
Our void-finder algorithm is an
improved and generalised version of the original algorithm proposed
by \citet{1998ApJ...497..534A}. The important feature of this algorithm is that it does not assume
a priori that voids are spherical and hence can be used to study the shapes of the voids.
We apply our void-finder algorithm to the Sloan Digital Sky Survey SDSS DR7 and build a catalogue of
voids. In parallel, we also apply our algorithm to a mock-SDSS DR7
catalogue, which we construct out of the Millennium I simulation.
The mock catalogue is given the same magnitude cut-off as SDSS DR7. In a
different version, we also set up a mock catalogue with the same
number density as SDSS, but a different magnitude cut-off.
This allows us to compare various properties of observed voids to
those predicted by $\Lambda$CDM and the semi-analytic model of
galaxy formation.

In Section \ref{sec:data}, we present our sample taken from the SDSS
DR7 catalogue. In Section \ref{sec:mock}, we present our mock catalogue.
In Section \ref{sec:algorithm}, we explain our void-finder
algorithm. In Section \ref{sec:obsmock}, we find the voids in the simulation and observation catalogues and discuss the numbers,
sizes, and shapes of the voids. 
In Section \ref{sec:abundance}, we study
the abundance of large voids in the observations and the mock catalogues. 
In Section \ref{sec:luminosity},
luminosities of voids as a function of their sizes are
presented and compared between the simulation and the observation. 
In Section \ref{sec:conclusion},  we conclude.

\begin{figure*}
\center{
\includegraphics[scale=0.6,bb=300 0 200 300]{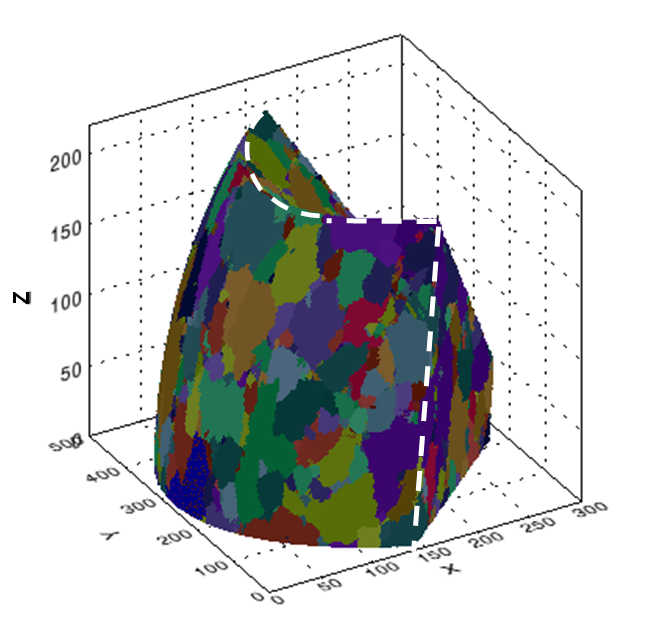}\hspace*{1cm}
\includegraphics[scale=0.75,bb=500 0 0 200]{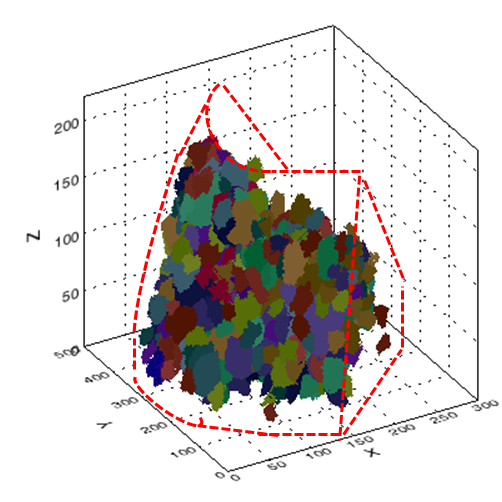}
\caption{ Right panel: initial voids in the observational data
  of SDSS DR7. Left panel: final voids after without small and edge voids.}}
\label{fig:voids}
\end{figure*}

\section{SDSS DR7: definition of the sample}
\label{sec:data}

We have selected the main galaxy sample of the seventh data release 
of the Sloan Digital Sky Survey (SDSS DR7) \citep{2009ApJS..182..543A}. 
The galaxy redshifts were corrected for the motion of the local group 
and are given in the CMB rest frame. 
The k-corrections for the SDSS galaxies were calculated using the
KCORRECT algorithm developed by \cite{2003AJ..125..227B} and \cite{2007AJ..133..734B}. 
The boundaries of our selected region of SDSS are: 
$135 < {\rm RA} < 235$ and $0 <  {\rm DEC} < 40$,  which contains
283076 galaxies. 
The choice of boundaries clearly is arbitrary. However, 
 the selected region in our study covers most
of SDSS DR7. We used spectroscopic data and applied a void algorithm
to volume-limited samples. Had we selected high-redshift galaxies, we would have had to consider
very bright galaxies (M < -21, -22), which would be meaningless for voids.      
All objects in this
selected region have a redshift error smaller than $2.5\times10^{-4}$
and the errors in their apparent ''Petrosian'' magnitudes of the r band, $m_r$,  are smaller than $0.1$.  
The absolute magnitudes of the galaxies were determined in the r band
using cosmological parameters; $H_0 = 100$ and the 
density parameters $\Omega_m=0.25$ and $\Omega_{\Lambda}=0.75$. 
Galaxies belonging to voids were identified
by using a volume-limited sample taken from the selected region. The
final subsample contains 68702 galaxies with
absolute magnitudes $ M_r < -19.9 $, which lie in the comoving distance interval
75-325 h$^{-1}$Mpc, corresponding to $0.02 < z < 0.12$.  

The selected region of the SDSS DR7 is shown in the left panel of Fig. 1.
The right panel of this figure shows the plot of 
the absolute r-band magnitude versus
comoving distance. The dark region in this plot illustrates the
selected volume-limited sample we used.

\section{Mock Millennium I catalogue: definition of the sample}
\label{sec:mock}

The Millennium I simulation was with a $N = 2160^3$  particles in a
comoving box of length $L = 500  h^{-1} {\rm Mpc}$ and 
mass resolution of $8.6 \times 10^8 h^{-1} M_{\odot}$. 
 The adopted cosmology is a $\Lambda$CDM model with $\Omega_m = 0.25$, 
$\Omega_b = 0.045, \Omega_{\Lambda}= 0.75, h = 0.73, n = 1$ and $\sigma_8 = 0.9$.
This value of $\sigma_8$ is higher than its present value of 
$0.8$ given by WMAP7 \citep{2011apjs...199..18K}, hence yielding more
power on larger scales. The evolution of baryons within these 
dark matter halos is predicted by different semi-analytic models.
Current semi-analytic models try to incorporate various complex processes such as gas cooling, reionization, 
star formation, supernova feedback, metal evolution, black hole growth, and active galactic nuclei (AGN) feedback 
({\it e.g.} \cite{2006\mnras.370.645B},\cite{2007MNRAS.375.2D},\cite{2011MNRAS.413..101G}).
Although the semi-analytic models are designed to match the
observational data as closely as possible, they can still fail in
certain aspects, for example the  low-mass galaxies with stellar-mass
($< 10^{9} M_{\odot}$) are slightly over-predicted. Consequently, to
remedy this problem, supernova feedback, a modified law for star
formation, or a different cosmological model are evoked (see {\it e.g.} \cite{2011MNRAS.413..101G};
 \cite{2012\mnras.308.510B}; \cite{2012\mnras.421.3450w}; 
\cite{2012MNRAS.421.2384M}). 

In this work, we used the mock galaxy redshift catalogue of the 
{Blaizot-ALLSky-PT-1} $^{1}$
 \footnotetext[1]{http://www.g\-vo.org/Millennium/Help?page=databases/
   mpamocks/blaizot2006\_allsky}, which was designed to 
mimic the SDSS and has an almost identical redshift distribution and a
very similar colour distribution. This mock catalogue 
was constructed by \cite{2005MNRAS.360..159B} using the mock map facility (MoMaF) code and the semi-analytic model presented in \cite{2007MNRAS.375.2D}.
Furthermore, to have a mock catalogue that resembles the SDSS DR7 galaxy survey as closely as possible, 
we selected a region in the simulation that lies in the same redshift range ($0.02 < z < 0.12$)
and has the same geometry.
Our mock volume-limited sample includes 68701 galaxies with stellar masses larger than $10^{9} M_{\odot}$ and  
brighter than $M_r < -20.16$,
roughly representing the galaxies brighter than $M_r < -19.9$ in the SDSS DR7 sample 
and covering a volume of  $1.2 \times 10^7 {\rm (Mpc/h)^3}$ in the volume-limited SDSS DR7. 
Consequently, the simulation sample has the same
galaxy number density  as the SDSS DR7 sample.

\begin{table}
\caption{ Characteristics of our volume-limited samples. }
\label{table:data} \footnotesize
\begin{tabular}{lll}
\toprule 
&\hspace*{-0.9cm}${\,{\rm Observation}}$ &\hspace*{-0.3cm}${\,{\rm Simulation}}$ \\
\hline
\hspace*{-0.25cm} ${\rm Sample\,\, Volume\,\, } ({\rm Mpc/h})^{3}$ & \hspace*{-0.5cm}$\approx 1.2\times 10^7$ & \hspace*{-0.2cm}$\approx 1.2\times 10^7$\\
\hspace*{-0.25cm} ${\rm Number\,\, of\,\, galaxies}$ & 68702 & 68701\\
\hspace*{-0.25cm} ${\rm Number\,\, of\,\, field\,\, galaxies}$ & 5873 & 5377 \\
\hspace*{-0.25cm} ${\rm Number\,\, of\,\, wall\,\, galaxies}$ & 62829 & 63324 \\
\hspace*{-0.25cm} ${\rm Number\,\, of\,\, void\,\, galaxies (field+faint)}$ & 26859 & 43666 \\
\hspace*{-0.2cm}${\rm Mean\,\, galaxy\,\, separation\,\,} ({\rm Mpc/h})$ & 6.22 & 6.35\\
 \bottomrule
\end{tabular}
\end{table}

\section{Void-finder algorithm}
\label{sec:algorithm}

Various definitions of voids have been suggested previously \citep{1981ApJ...248L..57K,1991MNRAS.248..313K,1994ApJ...431...20S,2003MNRAS.340..160B}
and a number of void-finding algorithms, some which assuming voids to be nearly
spherical, have been developed  (see {\it e.g.}  \citet{2002ApJ...566..641H}). 
We developed a method that does not assume a priori that voids are spherical, and is 
based on the original algorithm of \cite{1998ApJ...497..534A}. 
(Hereafter AM algorithm).  
The AM algorithm was originally written in 2D. We extended it to 3D
and adapted it for application to large datasets. The  algorithm does not 
constrain the voids to be of any
particular shape and hence can be used
to study the shapes of the voids and their deviations from sphericity.
 We emphasise that here we consider
the Aikio-Maehoenen statistics only as a tool for the relative measurement of some 
parameters of voids (eg. sphericity) in observational and simulated catalogues, 
and not as tool which would provide any absolute measurements.

Prior to applying of AM algorithm to 
our volume-limited galaxy sample, we
classified galaxies as wall or field galaxies. 
To distinguish between wall and field galaxies, we introduced the parameter
$d$, which is related to the mean distance of the third-nearest neighbour, $d_{3}$, 
and the standard deviation of its value, $\sigma$, by the following expression:
($d=d_{3}+1.5 \sigma$) \citep{2002ApJ...566..641H}. In our volume-limited galaxy
sample, all galaxies with a third-nearest 
neighbour distance, $d_{3}$, greater than this
selection parameter, $d$, were taken to be field galaxies and removed
from the galaxy sample. The remaining objects were identified as wall
galaxies. We remark that a field galaxy may lie within 
a void region, hence a void galaxy, whereas 
wall galaxies all lie in the cosmic filaments and
clusters and by definition are not to be found in voids. 

We found that the selection parameters, $d$, for observation
and simulation data are 5.96 and 6.16 Mpc/h, respectively, which means that 
$9\%$ of the galaxies in the observation and $8\%$ in the simulation 
are identified as field galaxies.
The details of the samples are given in Table 1.

To implement the AM algorithm,
the wall galaxies were gridded up in cells of size 1 Mpc/h. The AM algorithm starts on the Cartesian gridded
wall galaxy sample by defining a distance field (DF). For a given grid in a 3D galaxy sample the DF was defined as the distance to the nearest
particle. Then according to the value of DF for the closest neighbours of each
grid, the local maximum of the DF {\it subvoid} was calculated. To
assign each element in the grid sample to a subvoid, we employed the climbing
algorithm \citep{2001ApJ...546..609S} where for 
a unit cell bounded by the grid points, {\it i.e.} an elementary cell,  the gradient in DF to each
of the neighbouring cell is calculated. 
 In this method, the elementary cell and
every other cell along the climbing route is then assigned to a subvoid.
Finally, if the distance between two subvoids is less than both DFs, they will be
joined into a larger void. 

\begin{table*}
\caption{ Statistics of voids in the observation of SDSS DR7 and the mock simulation catalogue} 
\label{table:statistic} 
\begin{tabular}{|l c|cc|cc|cc|cc|}
\toprule
 & & \multicolumn{2}{c}{Observation} & & \multicolumn{2}{c|}{Simulation}  
\\
\cmidrule(r){3-4} \cmidrule(r){6-7}
 &  & \multicolumn{1}{c}{Number} & \multicolumn{1}{c|}{Volume ({\rm Mpc/h})$^{3}$ } 
& & \multicolumn{1}{c}{Number} & \multicolumn{1}{c|}{Volume ({\rm Mpc/h})$^{3}$} \\
\toprule
All voids &  &   4616 & 12541454  & &  4847 & 12555147   \\\hline
Edge voids  &  &   1148 & 7844214 $(62.5\%)$   & &   1193 & 7646672 $(61\%)$ \\
Small voids $(r_{eff}<7 {\rm Mpc/h})$ & & 3001 & 722062 $(5.8\%)$ & &  3085 & 845753 $(6.7\%)$\\
Voids in the final sample & & 467 & 3975178 $(31.7\%)$ & &  569 & 4062722 $(32.3\%)$\\
\bottomrule
\end{tabular}
\label{tabble:voidvolume}
\end{table*}

The {\it void volume} was estimated using the number of grid points inside a
given void multiplied by the volume associated with the grid cell. For each void, we
defined its {\it effective radius} ($r_{\rm eff}$) as the radius of a sphere whose volume is 
equal to that of the void. 

The configuration of each void in this
algorithm depends on the grid points, and subsequently  we
determined the {\it void centre} as the centre of mass identified by
the positions of the grid points that enclose an elementary cell.
 Following this standard method and giving the same weight to all elementary
cells, the centre of each void can be written as

\begin{equation} 
\ 
 X_{V}^{j} =1/N \sum_{i=1}^{N} x_i^j ,
\
\label{eq:center}
\end{equation}

\noindent
where $x_{i}^{j}$ $(j=1,2,3)$ are the 
locations of elementary cells and
$\it N$ is the number of cells in the void {$\it V$}. 
The {\it shape of a voids} is then characterised by
 the ratio of the total number of grid points, which lie
between its centre and its effective radius, 
to its volume. This ratio
is an indicator of the deviation of the void shape 
from sphericity. Ideally, for a
spherical void this ratio is equal to one. 

In the next section, we apply this algorithm to the SDSS DR7 and 
the mock catalogue to construct catalogues of
voids and study their characteristics.

\section{Voids in the SDSS DR7 redshift survey and in the mock catalogue}
\label{sec:obsmock}

\begin{table*}
\caption{ Sizes and sphericities of voids in the observation and
  simulation mock catalogues}
\label{table:voidstatistic} \footnotesize
\begin{tabular}{|l c|cc|cc|cc|cc|cc|cc|cc|cc|cc|cc|cc|cc|}
\toprule 

  & & \multicolumn{3}{c}{Effective radius $({\rm Mpc/h})$}&	& 
\multicolumn{3}{c}{Max-length $({\rm Mpc/h})$} & &
\multicolumn{3}{c}{Surface $({\rm Mpc/h})^{2}$} & &
\multicolumn{3}{c|}{Sphericity} \\\cmidrule(r){3-5} \cmidrule(r){7-9}\cmidrule(r){11-13} \cmidrule(r){15-17}
  & & \multicolumn{1}{c}{Max} & \multicolumn{1}{c}{Min} & \multicolumn{1}{c}{Median} & &
\multicolumn{1}{c}{Max} & \multicolumn{1}{c}{Min} & \multicolumn{1}{c}{Median}& &
\multicolumn{1}{c}{Max} & \multicolumn{1}{c}{Min} & \multicolumn{1}{c}{Median}& &
\multicolumn{1}{c}{Max} & \multicolumn{1}{c}{Min} & \multicolumn{1}{c|}{Median} \\\toprule
Observation& &30.47& 7.02&9.65&& 108.6&19.9&32.3&&35414&1214&2588&&0.82&0.22&0.71\\
  
Simulaion& &28.15& 7.00&9.08&& 103.1&19.1&30.1&&33018&1210&2276&&0.84&0.12&0.72\\

 \bottomrule
\end{tabular}
\end{table*}

We identified 4616 and 4847 voids of different sizes and shapes
in the SDSS DR7 survey
and in the mock catalogue, respectively.  We avoided problems due to boundary effects by
selecting voids that lie completely inside the geometrical
boundaries of our catalogues. Therefore, {\it edge voids}, those that touch the survey boundaries, are
removed from our void catalogue because of their under-estimated volumes and
distorted shapes (see Fig.2). 

The size of each void is characterised by its effective radius, defined in the previous section.
To avoid counting spurious voids, we set a threshold of 7 Mpc/h for the
minimum size of effective radii of voids in both samples. This threshold is
higher than mean distance between galaxies in the sample and helps 
to eliminate seemingly {\it small voids}
from the sample. 
After removing all spurious voids, we
had about 467
and 569 voids in our volume-limited sample of
the SDSS DR7 survey and the mock simulation data, respectively, which occupy $\sim 32\%$ of the
volumes of the samples.  In Table \ref{table:statistic},
we provide the void statistics. Hereafter all analyses are carried out on voids 
in the final sample, obtained after eliminating small and edge voids. 

\begin{figure*}
\center
{
\includegraphics[scale=0.3,bb=30 30 800 600]{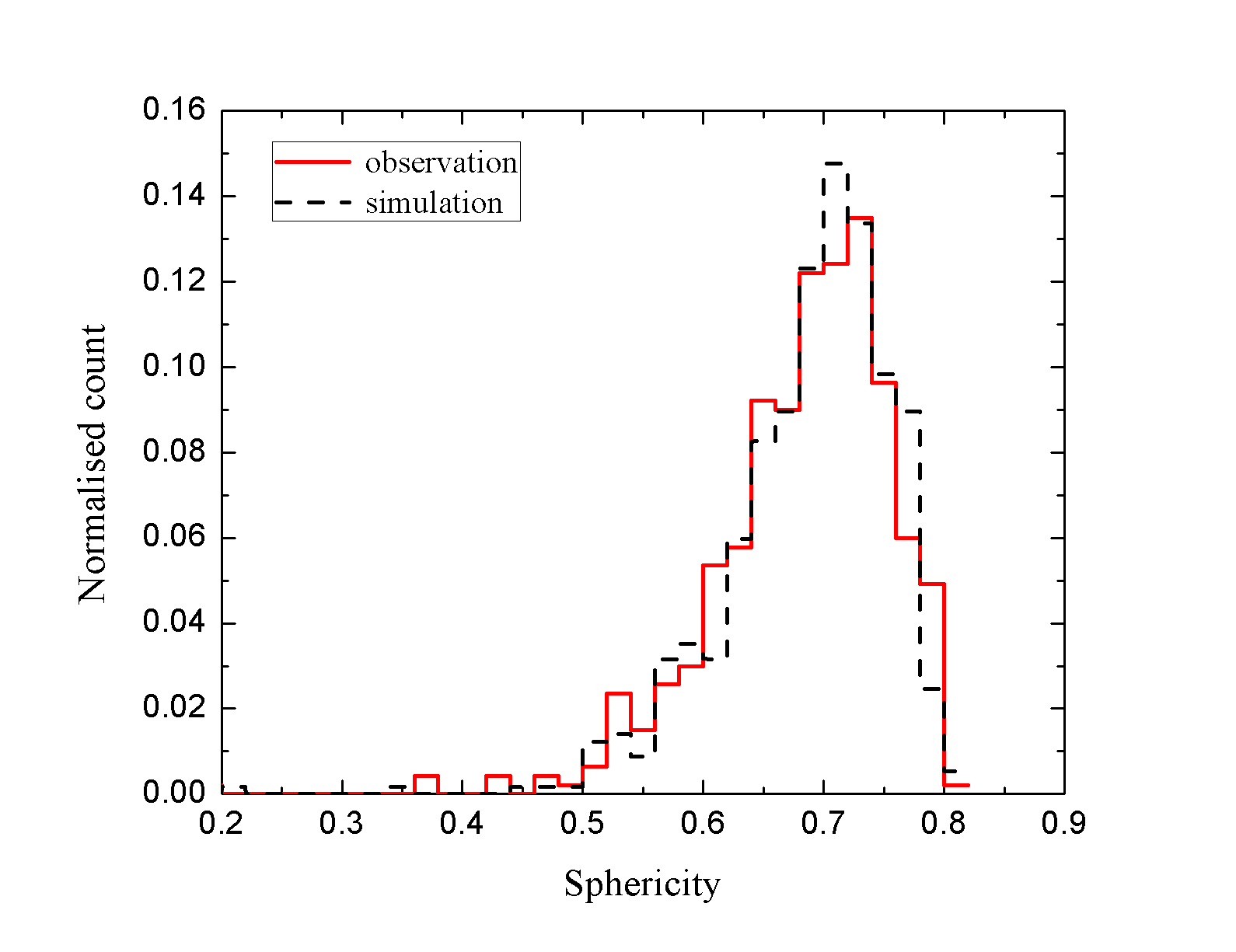}\hspace*{1cm}
\includegraphics[scale=0.3,bb=30 30 800 600]{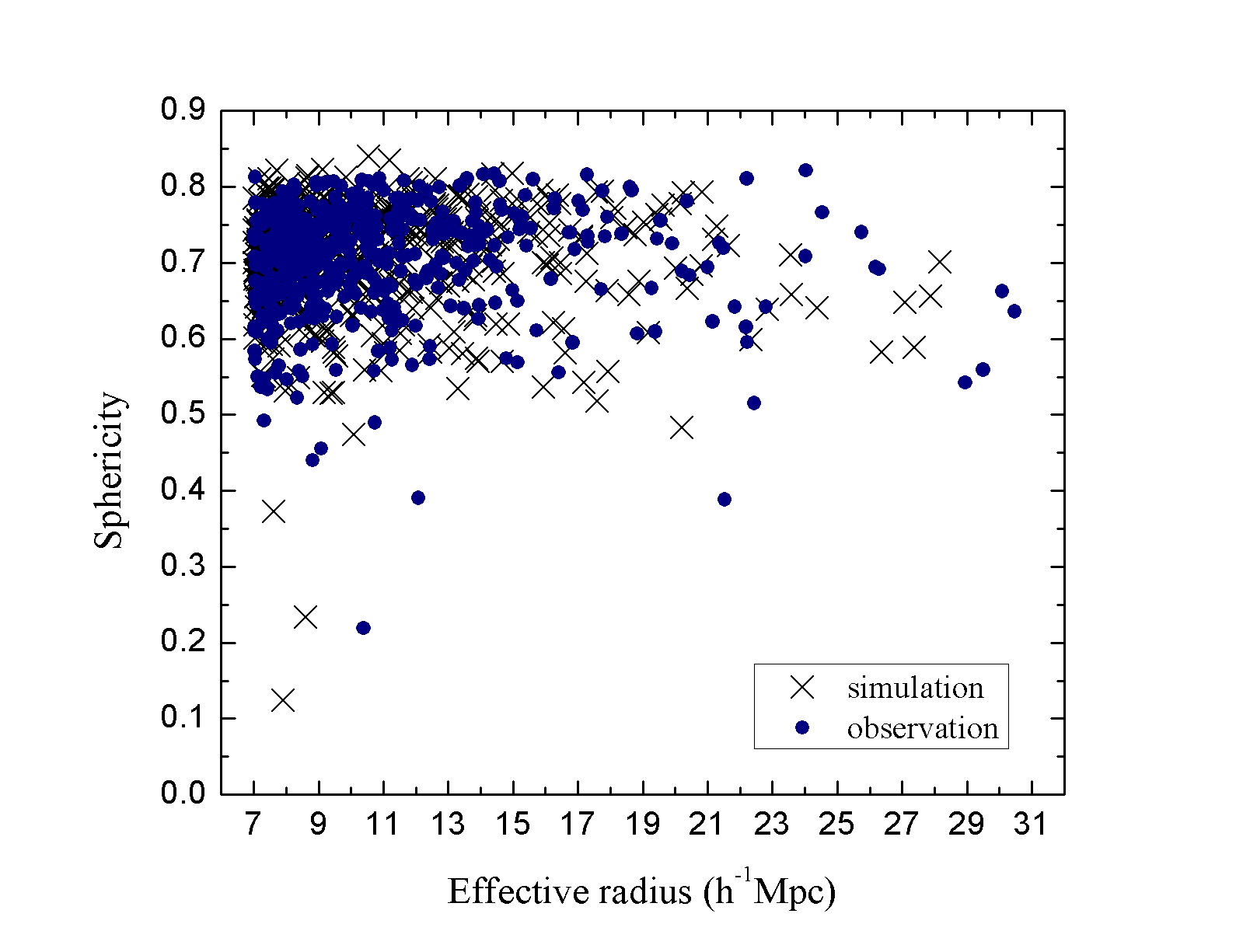}

} 
\caption{Left panel: distribution of sphericity is
  skewed towards larger sphericities, {\it i.e.} voids are mostly
  spherical. Right panel: plot of the sphericities versus
  the equivalent radii of the voids, demonstrating that voids
  become more spherical with increasing radii. There is no significant
  difference between the observation and the simulation and more data would be
needed to establish any disagreement between the two.}
\label{fig:sphericity}
\end{figure*}

Table 3 compares the statistical
properties of voids in the observed and mock catalogues. It shows
that the median of void sphericity in both samples is nearly $\sim 0.70$, which
indicates that voids tend to be mostly
spherical.  Fig.\ \ref{fig:sphericity} also shows that
voids tend to become more spherical with increasing radii. 
There is a good 
agreement between the mock catalogue and the SDSS observation, although
the observed voids seem to be marginally more 
spherical in general. More and better data are needed to 
see if the marginal difference reported here is of any significance.

\section{Abundance of large voids: the SDSS DR7 observation versus the mock catalogue} 
\label{sec:abundance}

\begin{figure*}
\center{
\includegraphics[scale=0.3,bb=30 30 800 600]{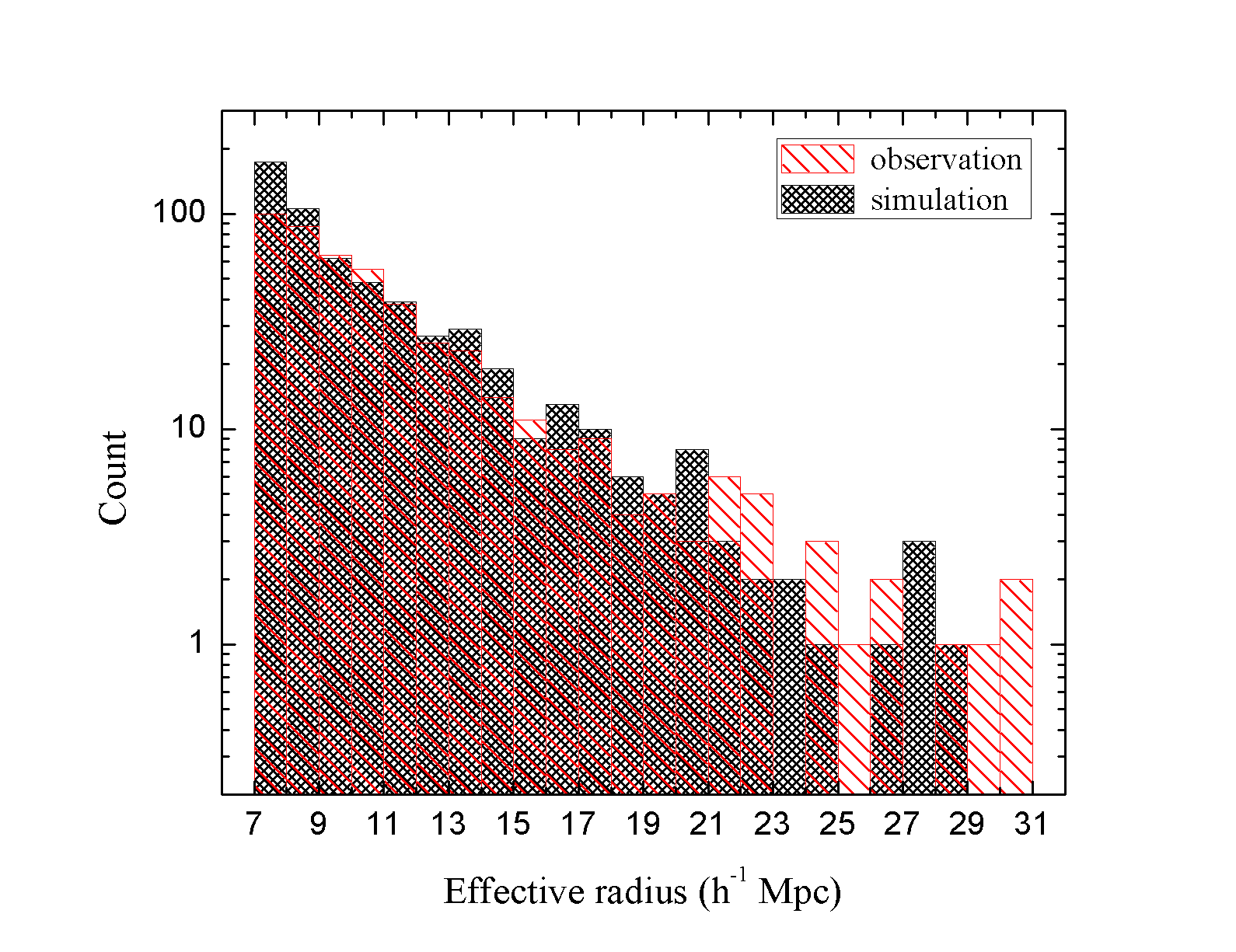}\hspace*{1cm}
\includegraphics[scale=0.3,bb=30 30 800 600]{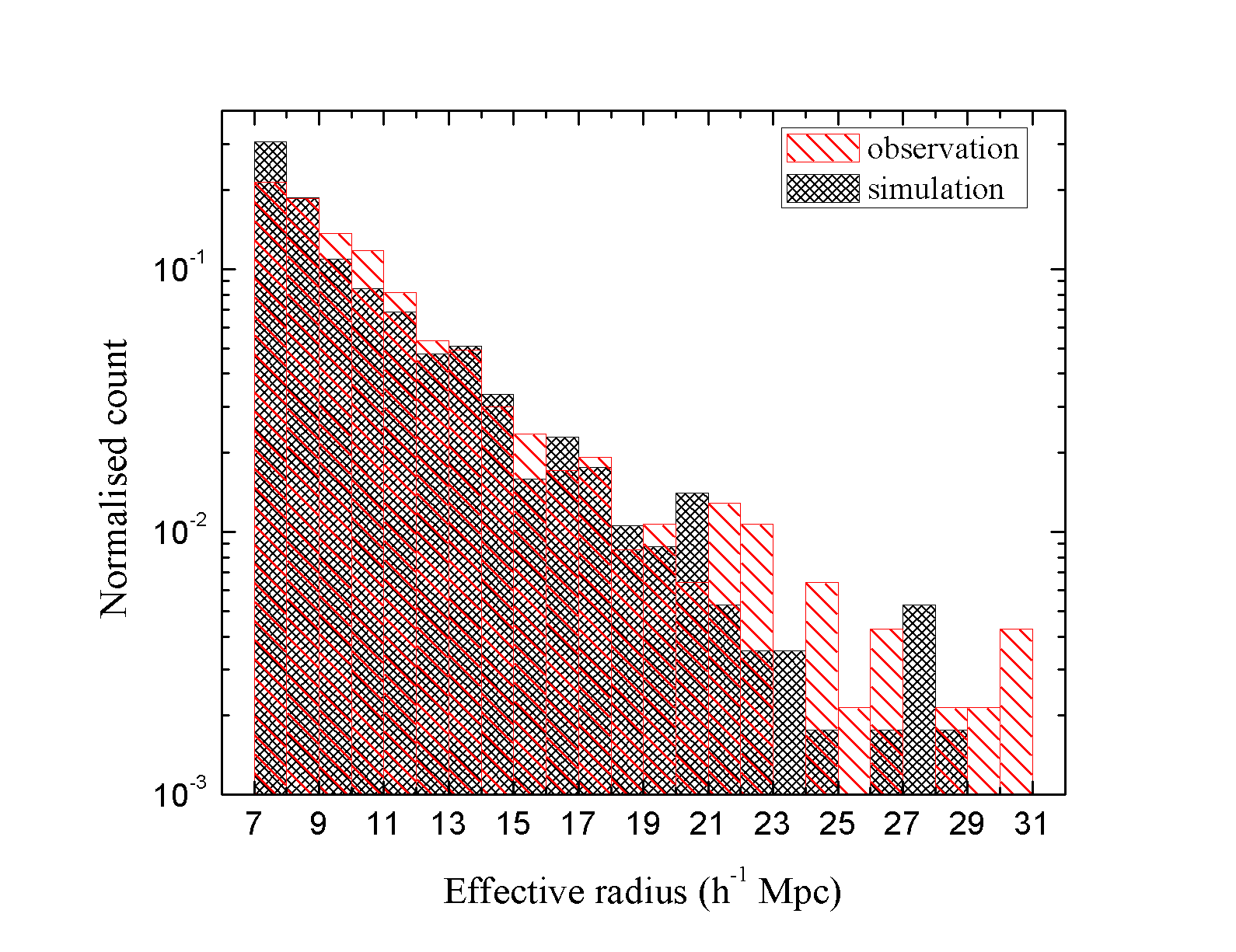}
\includegraphics[scale=0.3,bb=30 30 800 600]{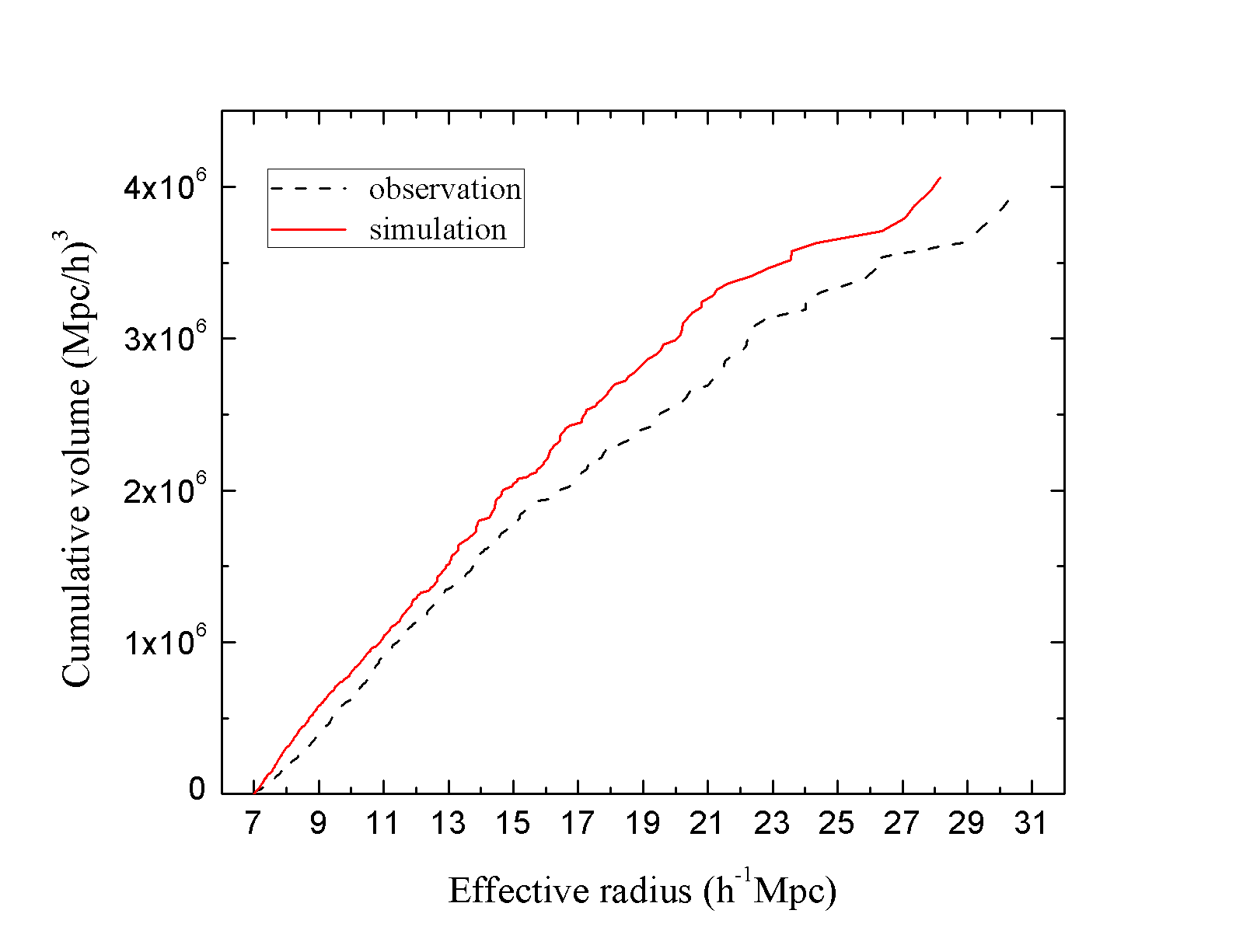}\hspace*{1cm}
\includegraphics[scale=0.3,bb=30 30 800 600]{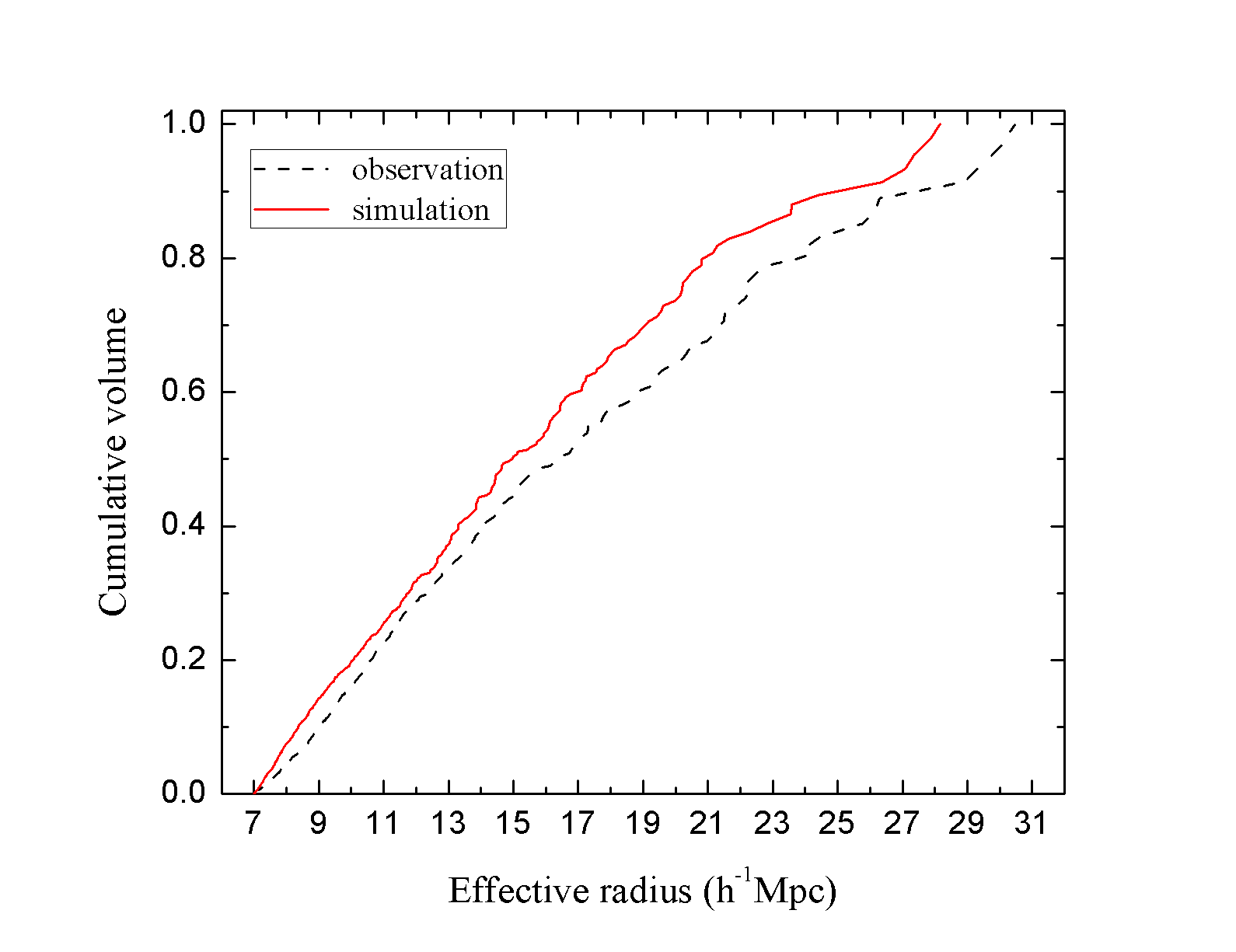}
} 
\caption{
 Top panel: Distribution of the void sizes in 
the observation and the simulation: larger voids are more
abundant in the observation.
 Bottom panel: Cumulative plots of the number of voids against their
 equivalent radii shows again that larger voids are more abundant in
 the observation.
The bottom plots show the volume/radius cumulative curves where both the
commulative volume and normalised volumes are plotted against
the effective radii of the voids. The histograms show that at large radii, there are more
 voids in the observation than in the simulation.  
 The lower panels demonstrate that the number and volume of voids are,
 in general, higher in the simulation
 than in the observation (see Table.2). Because there are only two catalogues, we cannot perform a proper error 
analysis and determine the error bars in these figures. 
However, we performed a Kolmogorov-Smirnov test that shows that the probability of the two samples
to have similar distributions is only about 0.004 and hence the difference between 
the two catalogues reported in these figures is statistically significant.
}
\label{fig:largevoids}
\end{figure*}

We compared the distribution of the void sizes in the
observation with the simulated mock catalogues.
Fig. \ref {fig:largevoids} shows that the volume occupied by voids is
larger
in the simulation than in the observation. In particular, both the histograms
and the commulative plots show that the largest voids are
absent from the simulation, whereas they are present in the observation.

The problem of large voids could be related to the over-abundance of
small galaxies, which would subsequently divide large voids into smaller ones. 
However, this could be resolved by proper biasing in modelling the galaxy formation and evolution.
Hence, the problem of large voids could be due to the shortcoming of the semi-analytic model of galaxy formation for the mock catalogue that we used here. 
A recent study that also compared the SDSS DR7 voids with those taken from a smoothed particle hydrodynamics (SPH) simulation and a halo-occupation model
and hence used a different model of galaxy evolution, seems to indicate
that the distribution of the void sizes agree in the two samples
\citep{2012MNRAS.421..926P}.  Hence, these void properties could be of potential importance in
distinguishing between different galaxy formation scenarios.

\section{Observed SDSS voids are less luminous than those in the mock catalogue}
\label{sec:luminosity}

\begin{figure*}
%\center
{
\includegraphics[scale=0.3,bb=30 30 800 600]{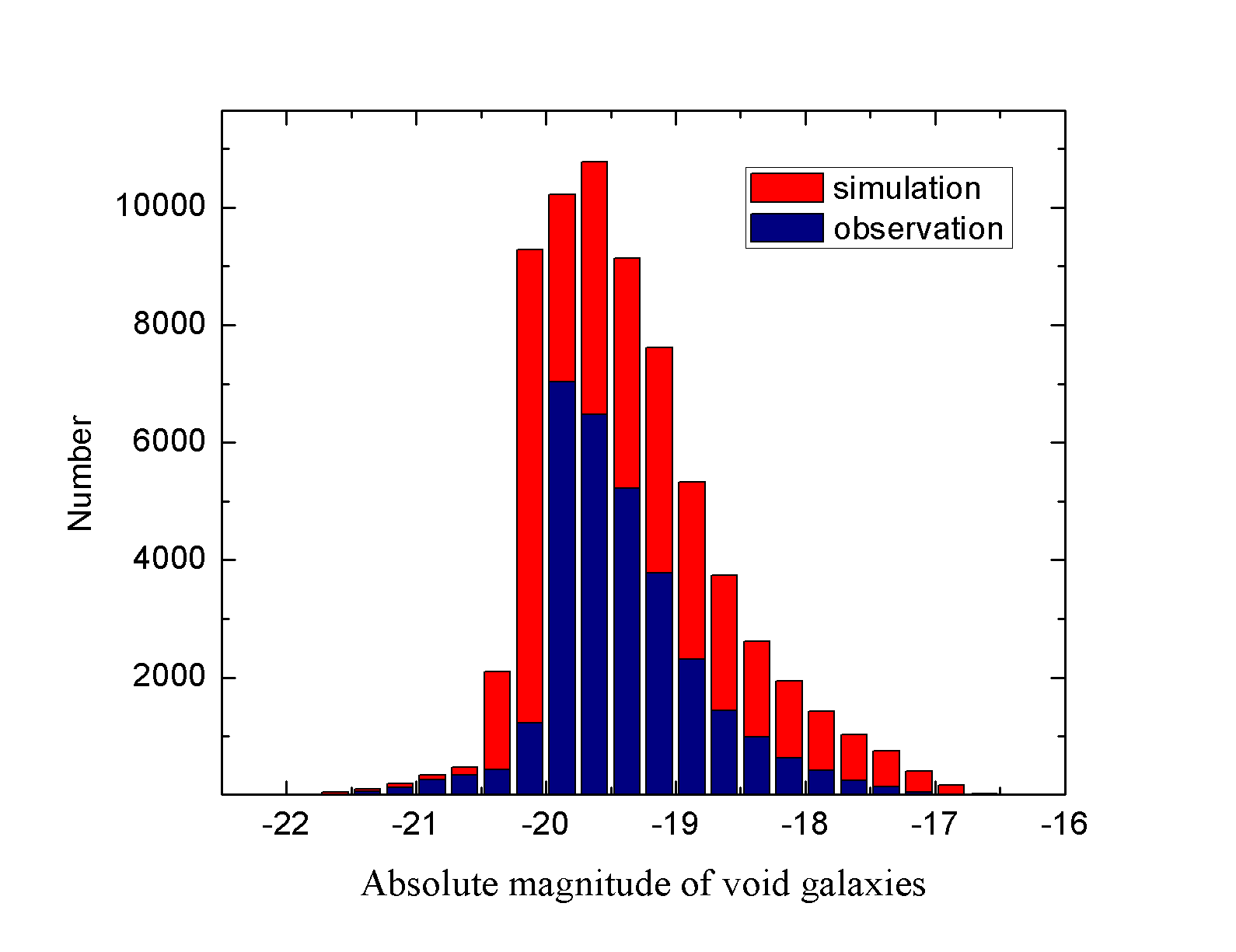}
}
\caption
{
Number of void galaxies plotted against their absolute magnitudes.
The luminosity range of void galaxies is nearly the same for the simulation and observation, which demonstrates that 
there is no bias imposed on the calculation of the void luminosities. Voids 
in the simulation contain more galaxies in almost all magnitude bands
and hence are more luminous than those in the observation. 
}
\label{fig:histo}
\end{figure*}

Prior to comparing the luminosities of the voids between simulation and observation, we checked that there was no bias between the two samples.
In Fig. \ref{fig:histo}, we plotted the  histogram of the absolute
magnitudes of field and faint galaxies that are found in the voids in the two catalogues. 
The figure shows that although there are more void galaxies in the mock catalogue 
than in the observation, the distributions are the same in both catalogues. Min and max
magnitudes are nearly the same, namely  $M\sim  -16.5$ in the
and M$\sim$ -22 in the observation and the simulation. 
This demonstrates that there is no bias between the two samples.

We comment that the void galaxies could be field galaxies or be field and faint galaxies.  We recall that the 
field galaxies are in the luminosity ranges
$M \textless -19.9$  in the observation and  $M \textless -20.16$ in
the simulation, but faint galaxies are less luminous than
these thresholds set in our  volume-limited sample (see Fig.\ref{fig:histo}). 
We stress again that to obtain the same number density in both samples, we have to consider
different luminosity thresholds in our two volume-limited samples (M=-19.9 \& M=-20.16). The
difference of luminosities is insignificant (about 0.26). Nonetheless, even if
we consider the same luminosity threshold for both samples ({\it e.g.} M=-19.9), we
derive the same result again and the galaxy luminosities in the simulation are higher
than in the observation.

\begin{figure*}
\center
{
\includegraphics[scale=0.3,bb=30 30 800 600]{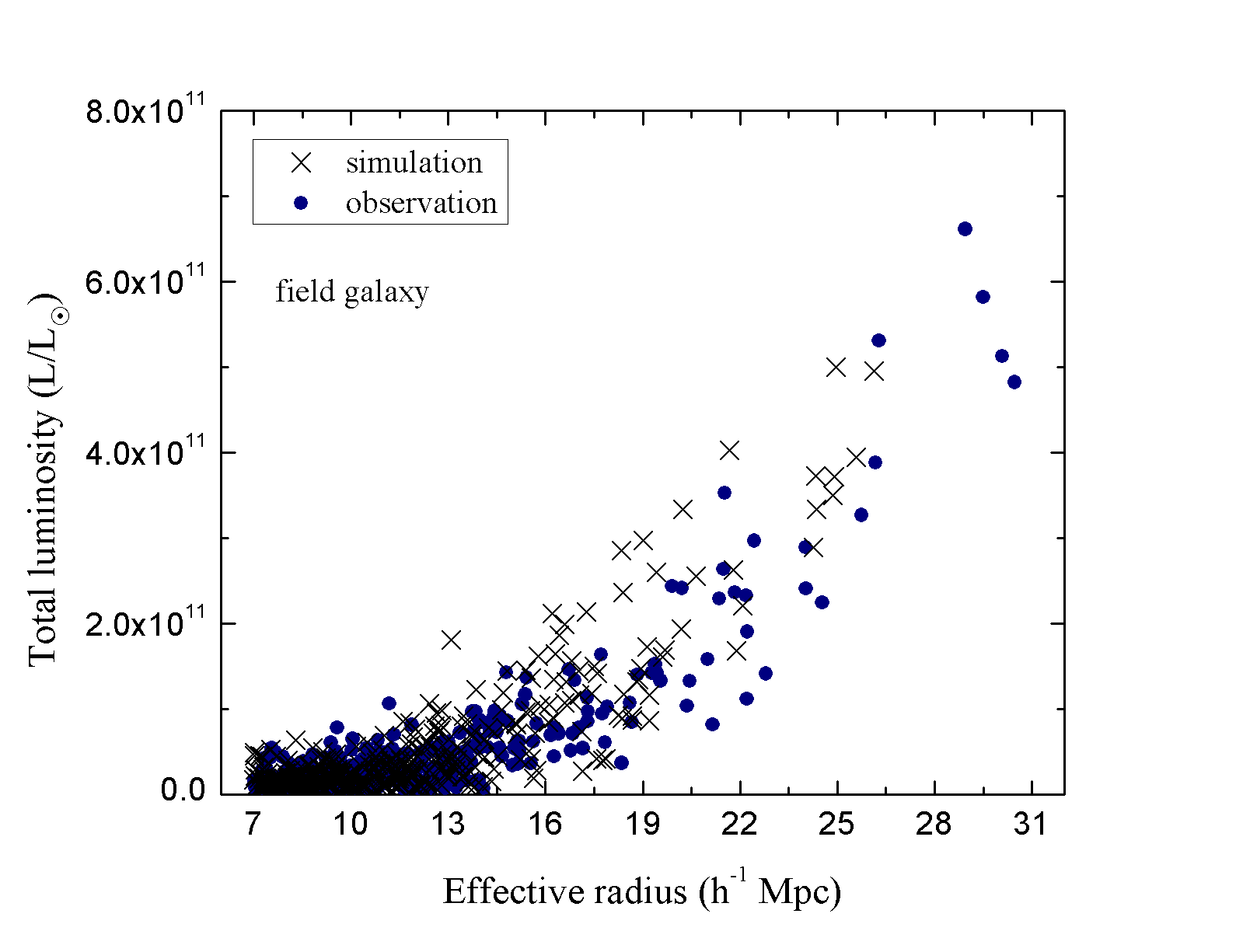}\hspace*{1cm}
\includegraphics[scale=0.3,bb=30 30 800 600]{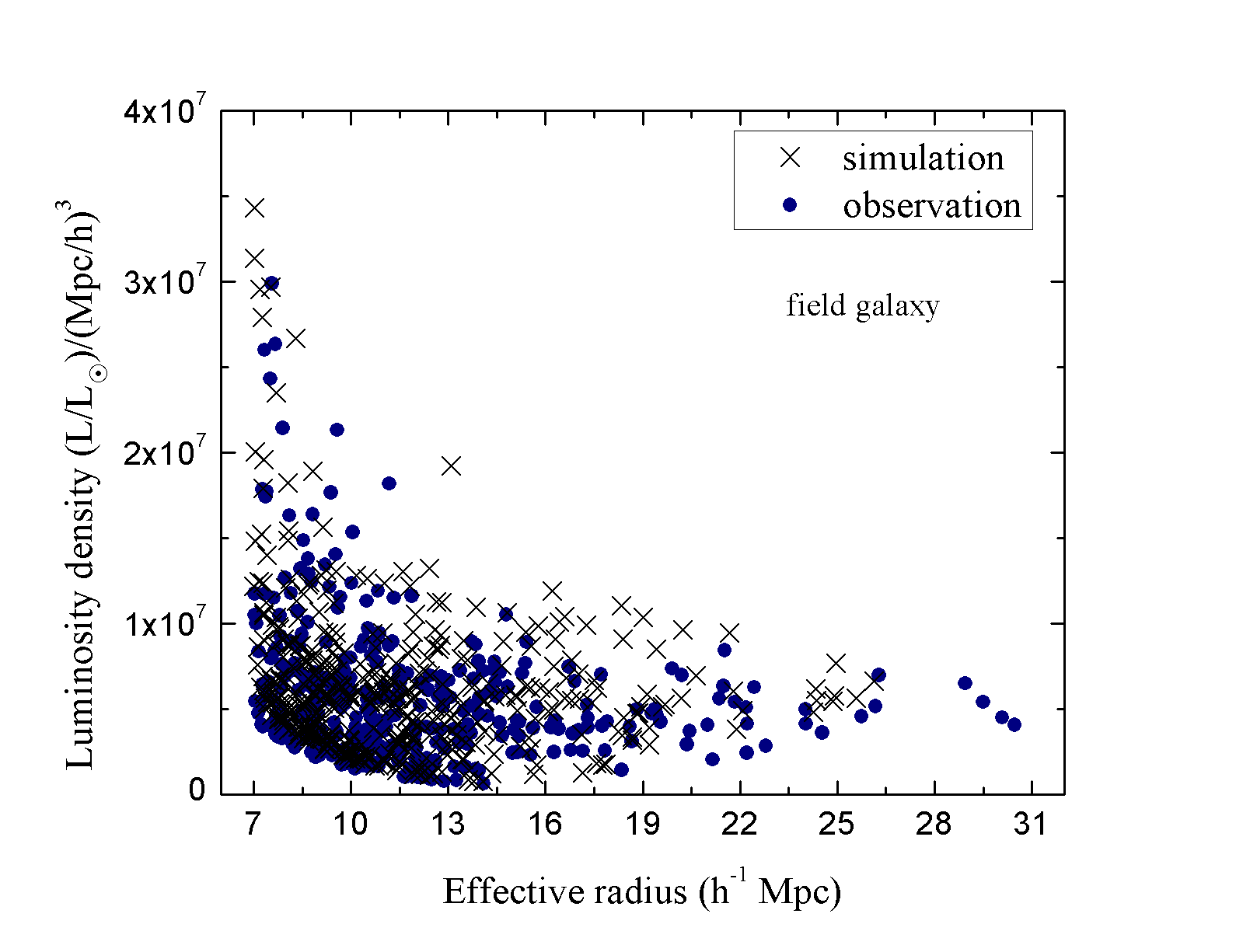}
\includegraphics[scale=0.3,bb=30 30 800 600]{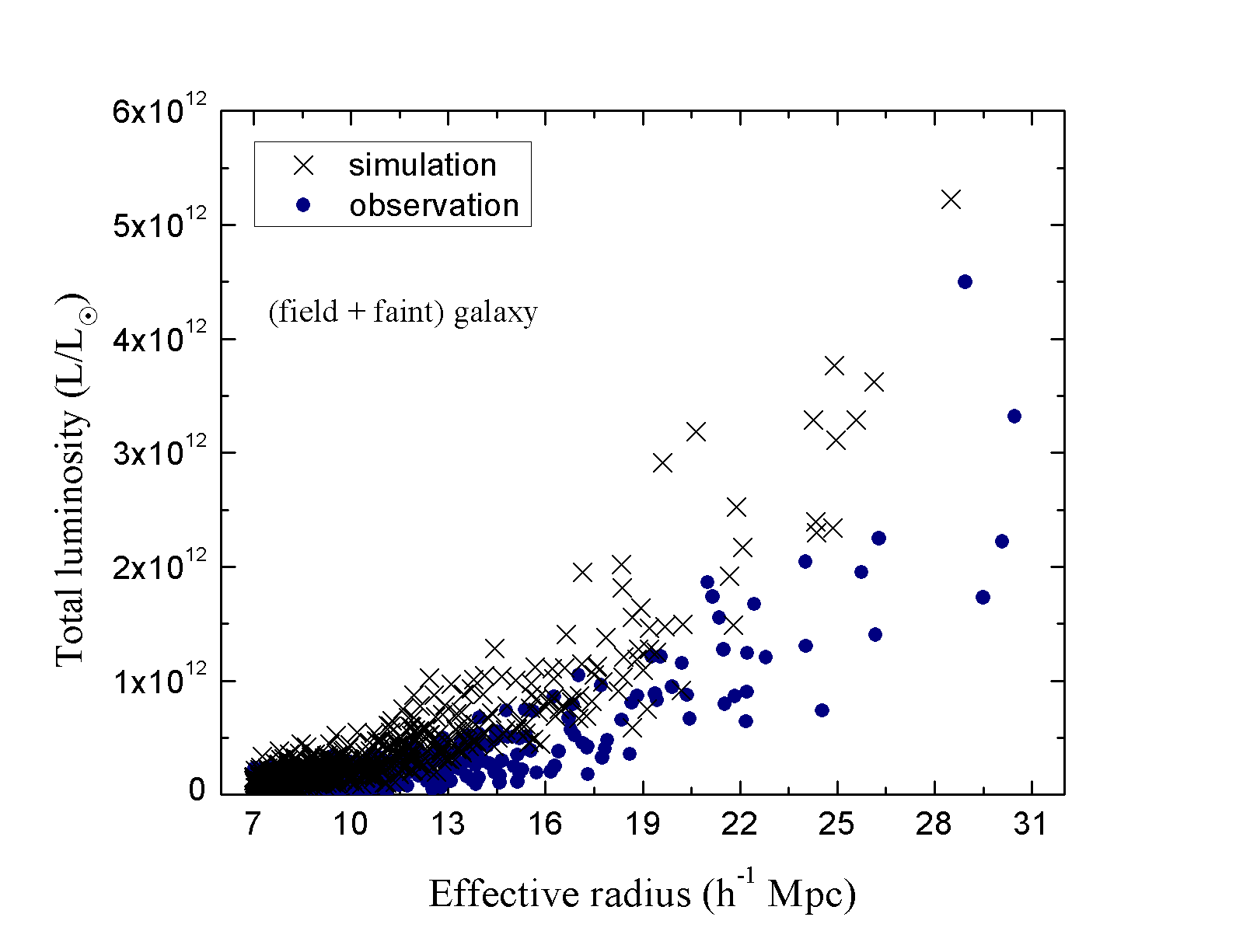}\hspace*{1cm}
\includegraphics[scale=0.3,bb=30 30 800 600]{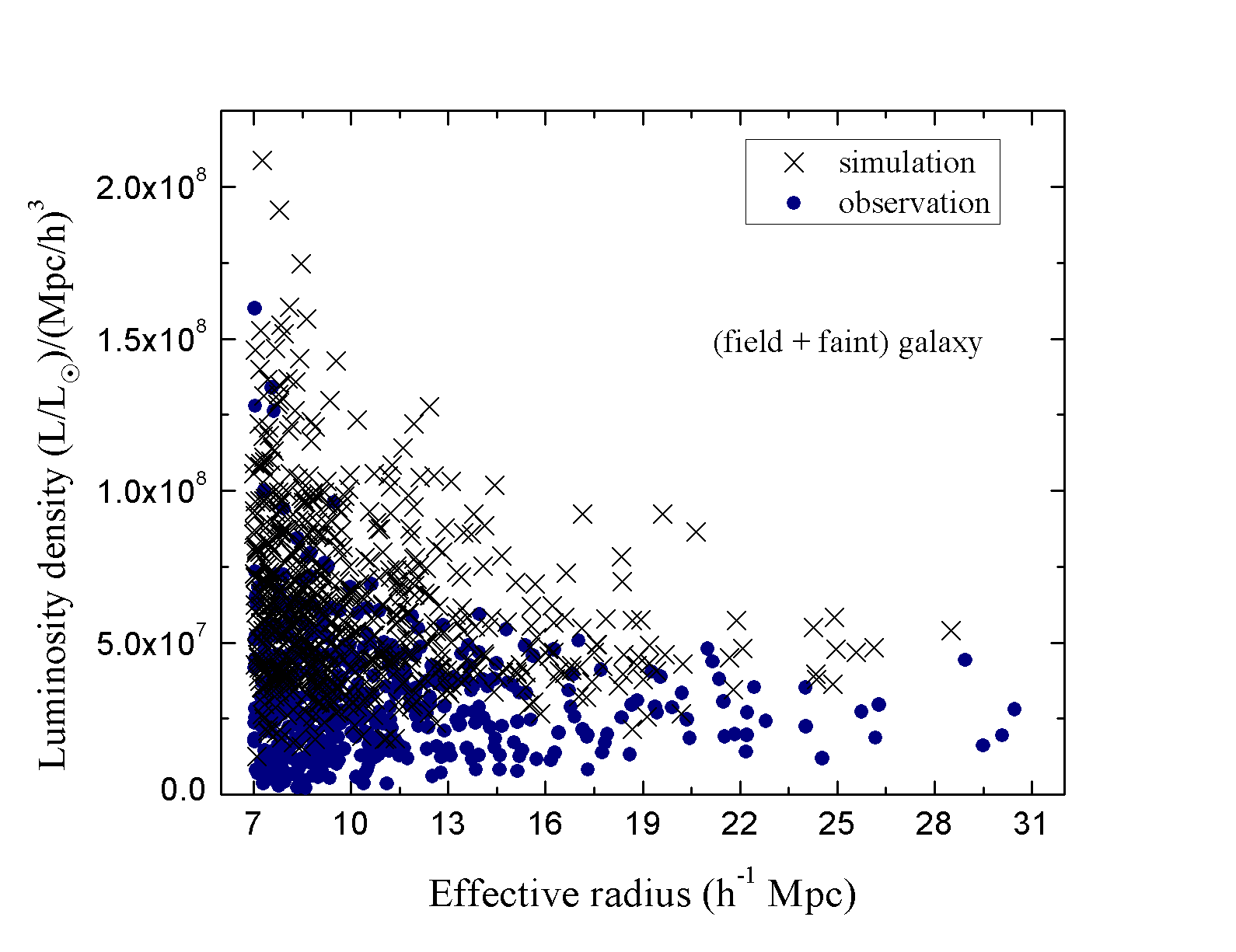}
} 
\caption{ 
{\it Top panel:} total luminosity and luminosity density of field galaxies 
ploted against the effective radii of the voids to which they
belong. Larger voids are less luminous in the observation than in the
simulation. This disagreement becomes more significant when faint objects are
also taken into account, as shown in the two plots of the lower panel.
Observed voids are clearly less luminous than simulated voids.
Note that the luminosity cutoffs are the same for the
observation and the simulation when faint galaxies are taken into account.
We expect this discrepancy to be even more significant
 than shown here because our Millennium I simulation uses a
 higher value of  $\sigma_8$ than given by WMAP7.  }

\label{fig:luminosity}
\end{figure*}

We compared the total luminosity of the voids and their luminosity per unit volume between the
observation and the simulation. The comparisons are shown in Fig. \ref{fig:luminosity}. 
The lower panel of Fig. \ref{fig:luminosity} shows that 
if we consider faint and field galaxies, large voids are clearly more
luminous in the mock catalogue than in the observation. 
However, the top panel of Fig. \ref{fig:luminosity} shows that  
if we consider only field galaxies, this discrepancy becomes less
prominent.
We emphasise that the lowest magnitude cut-off for both samples is nearly the
same when faint galaxies are considered (see Fig.\ref{fig:histo}). 
This discrepancy could be a sign of the over-abundance of small faint galaxies in the simulation.
The problem of empty voids could be related to the lacking large power of $\Lambda$CDM, even though 
the value of $\sigma_8$ used here is $0.9$, which is higher than its present value of 0.8 given by WMAP7. Hence,
this discrepancy is expected to be more significant for the WMAP7 value
of $\sigma_8$. 

\section{Conclusion }
\label{sec:conclusion}

We have carried out a parallel study of the voids in the SDSS
DR7 redshift survey and in a mock catalogue. The latter 
was extracted from the Millennium I simulation and
aims at replicating the observational biases and limitation of the SDSS DR7
catalogue. 

We found that  the total number and the volume occupied by the voids
are larger in the simulation than in the observation. We found 467 voids in SDSS
DR7 and 569 in the mock catalogue. The voids' pseudo-radii or effective radii ({\it i.e.} radii of an
equivalent spherical volume) range from 7 to 31 Mpc/h. The 
sphericities of voids also have similar distributions in the observation 
and the simulation. The voids also tend to
become more spherical with increasing effective radii.
Furthermore, large voids are less abundant in
the simulation and the mean void luminosities, as defined by the sum of the luminosities
of the galaxies they contain, is higher in the simulation. 
The aboundance problem of large voids could be related to the over-abundance problem of 
small haloes in $\Lambda$CDM ,which would then divide large voids into smaller ones in the simulation.
However, this problem is usually taken care of in models of galaxy
formation by suitable biasing or quenching of galaxy formation on
small scales.
The persistence of this problem could demonstrate that 
the semi-analytic model of galaxy formation used in the mock catalogue does not efficiently suppress
galaxy formation in small voids. Recent catalogues of voids in SDSS including also
the luminous red galaxies will be analysed in future works to obtain better statistics and shed
more light on this problem \citep{2012arXiv1207.2524S}.

We also found that voids are in general more luminous in the simulation than 
in the observation. This could be related to the lack of
power of $\Lambda$CDM on large scales.
 The value of $\sigma_8$ used in the Millennium I simulation is
$0.9$ compared to the value of 0.8 given by the WMAP7. 
The problem of  empty voids could then become even more
significant if the current value of $\sigma_8$ were used in the simulation.  
Hence, either the ingredients used in the semi-analytic model 
do not correctly reproduce the observations, or on a more fundamental level, the power spectrum of $\Lambda$CDM 
has too much power on small scales and too little on large scales, which cannot be remedied by realistic models of galaxy formation.

Acknowledgments:
We thank Sepehr Arbabi for collaborations and Habib Khosroshahi, 
Guilhem Lavaux, Gary Mamon, Joe Silk, and David Weinberg for discussions.
RM thanks IPM and ST and KV thank IAP for their hospitality.

%----------------------------------------------------

\end{document}